\documentclass[nonblindrev]{informs3} 

\OneAndAHalfSpacedXI

\usepackage{natbib}
 \bibpunct[, ]{(}{)}{,}{a}{}{,}%
 \def\bibfont{\fontsize{10}{12}\selectfont}%
\usepackage{multirow}
\usepackage{tabularray}
\usepackage{float}
\usepackage{graphicx}
\usepackage{codehigh}
\usepackage[normalem]{ulem}
\usepackage{booktabs}
\usepackage{pdflscape}
\usepackage{hyperref}

\TheoremsNumberedThrough     

\EquationsNumberedThrough    

\begin{document}



\RUNTITLE{AI Pose Analysis of ROM in Upper Body Resistance Training}

\TITLE{AI Pose Analysis and Kinematic Profiling of Range of Motion Variations in Resistance Training}

\ARTICLEAUTHORS{%
\AUTHOR{Adam Diamant}
\AFF{Schulich School of Business, York University, \EMAIL{adiamant@schulich.yorku.ca}}
}

\ABSTRACT{This study develops an AI-based pose estimation pipeline for quantifying movement kinematics in resistance training. Using videos from \citet{wolf2025lengthened}, comprising 303 recordings of 26 participants performing eight upper-body exercises under full (fROM) and lengthened partial (pROM) conditions, we extract joint-angle trajectories using five distinct deep-learning pose estimation models and a unified signal-processing framework. From these trajectories, we derive repetition-level metrics including range of motion (ROM) and  repetition duration. We use these outputs as dependent variables in a crossed random-effects model that accounts for participant-, exercise-, and model-level variability to assess systematic differences between ROM conditions. Results indicate that pROM reduces range of motion without significantly affecting repetition duration. Variance decomposition shows that pROM increases both between-participant and between-exercise variability, suggesting reduced consistency in execution. To enable cross-exercise comparison, we model ROM on a logarithmic scale and define \%ROM as the proportion of fROM achieved under pROM. While the estimated mean is $\sim56\%$, significant heterogeneity across exercises indicates that lengthened partials are not characterized by a fixed proportion of full ROM. The results demonstrate that AI-based motion analysis can provide reliable kinematic insights to inform evidence-based training recommendations.}


\maketitle
\section{Introduction}
A growing body of research in resistance training suggests that exercising at longer muscle lengths may confer distinct advantages for muscular adaptation. Specifically, training muscles in a lengthened position has been shown to promote hypertrophy \citep{maeo2023triceps, kassiano2023greater} and induce larger improvements in dynamic strength as compared to training in shortened positions \citep{pedrosa2022partial, pedrosa2023training, androulakis2024optimizing}. In fact, recent research suggests that performing exercises through a partial range of motion (pROM) in the lengthened portion of the movement can elicit adaptations that are comparable to those achieved using a full range of motion (fROM), which has been considered the gold standard for optimizing strength and hypertrophy \citep{newmire2018partial, kassiano2023roms, ottinger2023muscle, warneke2023physiology, wolf2023partial}. These findings have been amplified by the science-based lifting community, particularly through online forums and social media content \citep{RenaissancePeriodization,youtubeYouTube}. 

Despite the robust evidence of physiological adaption, translating these insights into practical recommendations remains nontrivial. Most studies on range of motion (ROM) in resistance training either measure performance across multiple predefined joint-angle intervals \citep[e.g.,][]{kassiano2023greater} or rely on qualitative labels such as “full” and “partial” repetitions \citep[e.g.,][]{wolf2025lengthened}. This introduces ambiguity regarding which joint angles constitute a partial repetition, whether those definitions are consistent across exercises with different strength curves (i.e., how a muscle’s force-producing capacity varies across joint angles and over an exercise’s range of motion), and whether the ROM achieved during training matches the intended prescription. Furthermore, most studies concentrate on a single exercise, potentially limiting generalizability even when precise joint-angle definitions are defined. Moreover, while many studies use randomized controlled trials to compare outcomes -- such as muscle thickness or strength -- across groups \citep[e.g.,][]{pedrosa2023training, varovic2025effects}, detailed aspects of \textit{movement execution} are often overlooked, including repetition duration, the relative durations of concentric and eccentric phases, and between-exercise and between-participant variability under different ROM conditions. As a result, the development of precise, execution-level training guidance and actionable coaching cues remain underexplored. Finally, the limited use of motion analysis -- largely due to the cost and logistical complexity of laboratory-based motion capture systems -- further constrains understanding of how kinematic factors contribute to the adaptation differences reported in the literature \citep[e.g.,][]{orangi2025effects}.

\begin{figure}[b]
    \centering
    \includegraphics[width=0.85\linewidth]{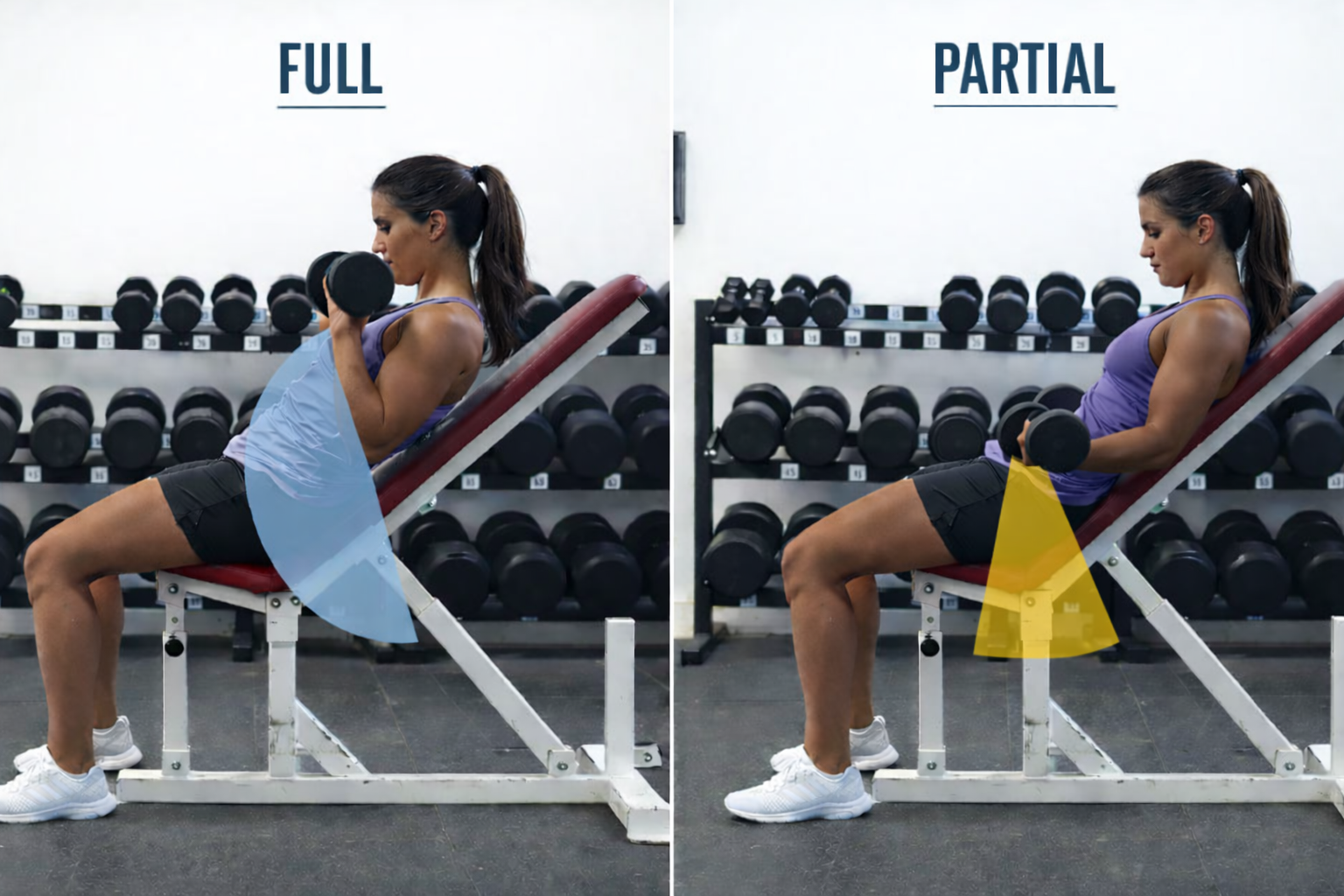} 
        \caption{Illustration comparing full (Left) versus partial range of motion in the lengthened position (Right) for the incline biceps curl. Image generated using the Dall-E 3 tool \citep{chatgpt_image_2026}.}
    \label{fig:visual_rom}
\end{figure}

To address these gaps, this study develops an AI-based pose estimation pipeline to quantify kinematics in upper-body resistance exercises performed under lengthened-partial (pROM) and full-range (fROM) conditions. Five methodologically distinct, deep-learning, human pose estimation models are implemented to ensure the robustness of downstream kinematic inferences. We describe the structure of these algorithms, the digital signal processing procedures used to extract reliable joint-level time series, detect exercise repetitions, compute repetition-level kinematic metrics, and assess the consistency of their outputs. The pipeline is applied to a dataset of videos from the publicly available repository associated with \citet{wolf2025lengthened}, which examined the effects of pROM and fROM training on muscular adaptation. In total, 303 videos covering eight resistance exercises and two ROM conditions were processed to extract joint-angle time series from which several metrics were derived, including the starting and ending angles of each repetition, per-repetition ROM (defined as the difference between maximum and minimum angles), repetition duration, concentric and eccentric phase durations, and the active body side (left or right). Figure~\ref{fig:visual_rom} illustrates the difference between ROM conditions using the incline biceps curl as an example.

Using the kinematic data extracted from these AI algorithms, we estimate a crossed random-effects regression model that captures participant-, exercise-, and model-level variation (random effects) as well as systematic differences associated with the execution of lengthened partials versus full ROM (fixed effect). This framework allows us to address three research questions. First, we quantify differences in repetition and phase durations between pROM and fROM. Second, we conduct a variance decomposition to determine how the primary sources of variability are affected by ROM condition. Third, we examine whether an exercise-independent criterion for defining pROM exists across upper-body exercises, thereby evaluating whether an “ideal” ROM proportion characterizes lengthened partial training. In doing so, this study extends prior work, demonstrating agreement between motion-capture systems and computer vision models \citep[e.g.,][]{moro2022markerless,carriere2025can}, by applying pose-based tracking to characterize resistance training execution.

The statistical results indicate that lengthened partials do not represent a consistent fraction of total angular displacement. Rather, they are better characterized as an exercise-dependent construct. In addition, despite the reduced range of motion, pROM repetitions exhibit similar durations to fROM, indicating that pROM affects the magnitude of mechanical excursion but not the temporal structure of movement execution. Variance decomposition further shows that pROM increases both between-participant and between-exercise variability relative to fROM. This greater heterogeneity suggests that partial-range movements may require clearer instructional guidance to maintain repetition consistency. Together, these findings complement existing hypertrophy and strength studies by showing that lengthened partials exhibit distinct kinematic characteristics.

In Section~\ref{sec:methods}, we describe the dataset, the AI-based pose estimation pipeline, and the crossed random-effects model used to compare pROM and fROM. Section~\ref{sec:results} presents the results for repetition duration and range of motion, reports the variance decomposition, and introduces an empirical specification to assess whether a standardized definition of lengthened partials exists. Section~\ref{sec:conclusion} summarizes the findings, discusses their implications, and outlines directions for future research.

\section{Data and Methodology} \label{sec:methods}
In this section, we describe the video data, the AI-based processing pipeline, the construction of the final tabular dataset, and the crossed random-effects model used for statistical inference.

\subsection{Training Videos}
Videos were obtained from the publicly available data repository (\href{https://osf.io/a6cpz/overview}{https://osf.io/a6cpz/overview}) accompanying \citet{wolf2025lengthened}, which investigated the effect of lengthened partial (pROM) versus full ROM (fROM) resistance training across eight upper-body exercises in 26 participants using a within-subject design. In the study, each participant performed a series of upper-body exercises under both ROM conditions during an 8-week intervention period. Training was conducted twice weekly, comprising four exercises per session and four sets per exercise, and was under the supervision of at least one research assistant at all times. Study participants performed all sets to momentary muscular failure, defined as the inability to complete another repetition despite maximal effort, while research assistants provided verbal encouragement and monitored adherence to the prescribed ROM. The order of training was randomized across sessions, loads were adjusted to maintain the target repetition range and intensity for all ROM conditions, and multiple repetition ranges were used across exercises. All training sessions were separated by at least 48 hours. Pre- and post-intervention testing assessed muscle thickness \citep{coleman2024gaining} and upper-body strength endurance via 10RM testing to evaluate muscular hypertrophy and strength adaptations.

Participants were eligible for inclusion in the study if they were 18–40 years old, free of cardiorespiratory or musculoskeletal disorders, engaged in regular upper-body resistance training ($\geq$1 session/week for $>$80\% of the prior 6 months), and reported no anabolic steroid use in the past year. Then, over the study period, principal investigators recorded each participant once per ROM condition and exercise modality to demonstrate an ecologically valid implementation of the fROM and pROM protocols. Videos captured a single set per participant, with no standardization of camera angles. However, not all combinations of exercises and ROM conditions were available in the repository. Specifically, of the 309 videos that were recorded\footnote{If every participant completed all exercises under both ROM conditions, the total number of videos would be 416.}, six were duplicates\footnote{Three duplicates for Participant 21, two for Participant 18, and one for Participant 24.}. 

\subsection{Pose Estimation and Joint-Angle Time Series Analysis} \label{sec:pose_pipeline}
The videos were processed using five methodologically distinct deep learning–based human pose estimation models: 
\textit{RTMPose} \citep{jiang2023rtmpose},  \textit{DWPose} \citep{yang2023effective}, \textit{RTMO} \citep{lu2023rtmo}, \textit{MediaPipe Landmarker} \citep{lugaresi2019mediapipe}, and \textit{YOLO26} \citep{yolo26_ultralytics}. \textit{RTMPose} and \textit{DWPose} follow a top-down approach in which a person is first localized in the video using an object detector \citep{redmon2018yolov3}, after which joint coordinates are estimated in the detected regions. Within this paradigm, the models differ in their underlying feature extraction architectures. Specifically, \textit{RTMPose} employs a convolutional backbone and was configured to produce 3D pose estimates. 
\textit{DWPose} performs 2D pose estimation using a high-resolution convolutional backbone with specialized decoding heads to predict whole-body keypoints. In contrast, \textit{RTMO} adopts a bottom-up approach, predicting joint candidates and their spatial associations across the entire image before assembling them into individual skeletons. \textit{YOLO26} uses a single-stage, detect-and-estimate architecture that jointly predicts bounding boxes and body keypoints within a unified neural network. Finally, \textit{MediaPipe Landmarker} employs a regression-based approach that predicts landmark coordinates from image features and is optimized for real-time inference. All models operated on RGB video at a resolution of $1280 \times 1280$ without subject-specific calibration or manual annotation and were implemented using Pytorch \citep{paszke2019pytorch}.

From the extracted 2D/3D joint coordinates, a time series of joint angles was computed. That is, for frame $t$, the angle $\theta_t$ of a given joint was calculated from three anatomically adjacent landmarks: a proximal joint $(\mathbf{a}_t)$, the joint center $(\mathbf{b}_t)$, and a distal joint $(\mathbf{c}_t)$. Specifically, 
\begin{equation*}
\theta_t = \cos^{-1}\!\left(
\frac{(\mathbf{a}_t - \mathbf{b}_t) \cdot (\mathbf{c}_t - \mathbf{b}_t)}
{\|\mathbf{a}_t - \mathbf{b}_t\| \; \|\mathbf{c}_t - \mathbf{b}_t\|}
\right).
\end{equation*}

For each exercise, a single relevant joint angle was selected to characterize ROM. The elbow flexion–extension angle was used for pressing, curling, rowing, and triceps-extension movements, while the shoulder angle was used for the Lat Pulldown. These choices were guided by biomechanical relevance to the primary torque-producing joint and by the robustness of the landmark detection algorithms. Further, elbow/shoulder landmarks were consistently visible across recordings and minimally affected by occlusion or camera perspective, resulting in stabler angle trajectories.

Exercising side (left vs. right) and ROM (partial vs. full) condition was determined using externally recorded labels that were merged with the pose estimation outputs. Joint-angle trajectories were first filtered using the landmark confidence scores provided by each model; frames with values below a threshold were treated as missing. The time series were then re-indexed to a continuous frame sequence and missing values were reconstructed using piecewise cubic Hermite interpolation \citep{lu2018novel}. Outliers were identified using a rolling median absolute deviation filter and replaced with missing values when z-scores exceeded the 95th percentile  \citep{yaro2023outlier}. The interpolated trajectory was subsequently smoothed using one of several filtering pipelines. Specifically, depending on the configuration, the signal was processed using (i) Savitzky–Golay \citep{schafer2011savitzky} filtering alone; (ii) median filtering followed by Savitzky–Golay filtering; or (iii) a frequency-domain low-pass filter \citep{cerna2000fundamentals} based on the Fast Fourier Transform (FFT). The smoothing method and its associated parameters were selected during hyperparameter optimization.

Exercise repetitions were identified using a trough-based segmentation approach. The joint-angle trajectory was first oriented so that peaks corresponded to the end of the concentric phase. A robust estimate of movement amplitude was then computed as the difference between the 5th and 95th percentiles of the joint-angle trajectory. The dominant repetition period was estimated using complementary spectral and temporal analyses. That is, an initial cadence estimate was obtained from the FFT spectrum by identifying the dominant frequency within plausible bounds. A second estimate was derived using the autocorrelation function. When the two estimates were sufficiently similar, a weighted average was used. Otherwise, the estimate with the stronger periodicity score was selected. 
Peaks and troughs were detected using a prominence-based algorithm, and extrema closer than a minimum spacing were suppressed. Additional amplitude-based hysteresis constraints ensured that oscillations caused by noise were not interpreted as a valid repetition.

Candidate repetitions were defined as consecutive trough–peak–trough triplets. Several filters were then applied to reject implausible measurements. First, repetitions were required to exceed a minimum absolute ROM threshold. Second, repetition durations had to fall within physiologically plausible bounds. Third, amplitude around the peak had to exceed a minimum fraction of the overall movement amplitude. Finally, validation checks constrained repetitions based on the ratio of concentric to eccentric phase durations and their consistency with the estimated repetition period. 

After candidate repetitions were identified, a relative ROM filter was applied to remove shallow movements. Repetitions whose ROM fell below a specified proportion of the median were discarded. To minimize bias arising from initial setup and terminal fatigue, the first and last repetitions of each set were excluded from downstream statistical analyses. For each retained repetition, ROM was defined as the angular difference between the peak and nadir contraction angles. Repetition duration was calculated as the time between successive troughs where  the concentric and eccentric phases were separately segmented. In addition, several quality-control metrics were computed for each video. These included the average landmark confidence score, a signal-to-noise (snr) ratio derived from the variance of the smoothed joint-angle trajectories relative to the residual noise, the number of detected repetitions, and the coefficient of variation of repetition durations. Videos were excluded when they exhibited low average landmark confidence, insufficient signal-to-noise ratio, an insufficient number of detected repetitions, or highly irregular repetition cadence.

Parameters governing smoothing, repetition detection, and quality-control thresholds were determined using automated hyperparameter optimization with the Optuna package \citep{akiba2019optuna}. The  procedure jointly tuned the digital signal-processing parameters and exclusion criteria to minimize discrepancies between automatically detected and manually annotated repetition counts while maintaining a high inclusion rate of usable videos. Hyperparameter tuning was performed separately for each pose estimation model to account for differences in keypoint accuracy and noise characteristics. For each analyzed video, summary kinematic variables were extracted from the retained repetitions, including the mean and standard deviation of ROM, repetition duration, and concentric and eccentric phase durations, as well as the average starting and ending joint angles.

\subsection{Statistical Analysis} \label{sec:stats}
Given the within-subject design of the original experiment, we leveraged the joint-angle time series for statistical analysis. However, not all participants completed every exercise under both pROM and fROM, and the number of successfully segmented repetitions varied by exercise and ROM condition, resulting in an unbalanced and crossed dataset. In addition, each video was processed using multiple pose estimation pipelines, producing several algorithmic estimates of the same kinematic quantity. To account for these sources of variation, we fit a crossed random-effects regression model \citep{baayen2008mixed} with participants, exercises, and pose estimation pipelines as random factors.

In this specification, participant $i$ contributes kinematic score $Y_{iegm}$, an estimated standard deviation $s_{iegm}$, and number of repetitions $k_{iegm}$, for exercise $e$, ROM condition $g \in \{\textrm{pROM},\textrm{fROM}\}$, and pose estimation pipeline $m$. Because participants, exercises, and pose estimation models may each be viewed as samples from broader populations, and each study participant completed multiple exercises under both ROM conditions, we estimated the following random-effects model:
\begin{equation} \label{meta_regress}
Y_{iegm} = \beta_0 + \beta_1\,\mathbf{1}_{\{g=\mathrm{pROM}\}} + \beta_2\,\mathbf{1}_{\{\mathrm{sex}_i=\mathrm{F}\}} 
+ p_i + u_e + r_m + (q_i + v_e)\,\mathbf{1}_{\{g=\mathrm{pROM}\}} + \varepsilon_{iegm},
\end{equation}
where $\beta_0$ denotes a common intercept, $\beta_1$ represents the average fixed effect of the pROM condition relative to fROM, and $\beta_2$ is the fixed effect of sex noting that only four participants were female. The random terms $p_i$ and $q_i$ correspond to participant-specific intercept and slope effects, respectively, capturing between-participant heterogeneity in both baseline performance and in their response to the pROM condition. Similarly, $u_e$ and $v_e$ represent exercise-specific random intercept and slope effects, which account for systematic differences across exercises and their differential sensitivity to pROM. Finally, $r_m$ denotes the random intercept associated with the pose estimation pipeline, capturing systematic differences across model architectures equally affecting pROM and fROM measurements. The random effects are assumed to follow normal distributions:
\begin{equation*}
\begin{bmatrix} p_i\\ q_i \end{bmatrix}
\sim \mathcal{N}\!\left(\mathbf{0},
\begin{bmatrix}
\tau_p^2 & 0 \\
0 & \tau_q^2
\end{bmatrix}\right),
\quad
\begin{bmatrix} u_e\\ v_e \end{bmatrix}
\sim \mathcal{N}\!\left(\mathbf{0},
\begin{bmatrix}
\tau_u^2 & 0\\
0 & \tau_v^2
\end{bmatrix}
\right),
\quad
r_m  \sim \mathcal{N}(0, \tau_r^2),
\end{equation*}
where $\tau_p^2$, $\tau_u^2$, $\tau_r^2$ denote the variance components for the random intercepts at the participant, exercise, and pose estimation pipeline levels, respectively, while $\tau_q^2$ and $\tau_v^2$ denote the corresponding variance components for the random slopes associated with the pROM condition.

The residual term $\varepsilon_{iegm}\sim\mathcal{N}(0,\sigma^2_{iegm})$ captures observation-level error, with sampling variance $\sigma^2_{iegm}=s^2_{iegm}/k_{iegm}$ which is used to define inverse-variance weights $1/\sigma^2_{iegm}$. Model parameters in the crossed random-effects regression specification \eqref{meta_regress} were estimated by restricted maximum likelihood (REML) using the \texttt{metafor} package \citep{viechtbauer2010conducting} in the \texttt{R} programming environment \citep{rprogramming} to account for the unequal variance weighting. Unless otherwise noted, we report model-based REML standard errors for all fixed-effect parameter estimates.

Conceptually, the fixed-effect terms quantify the overall difference between ROM conditions after adjusting for sex, which has been previously shown to influence performance outcomes \citep[e.g.,][]{mc2021, Nuzzo_2022}. The participant-level random intercepts and slopes capture individual variability in both baseline performance and responsiveness to the pROM condition, reflecting that participants may differ not only in their innate abilities but also in how their performance changes due to the treatment. Similarly, exercise-level random intercepts and slopes allow the magnitude of the pROM effect to vary across exercises, reflecting the diversity of upper-body movements that may be included in a training program \citep{jefitExerciseDatabase}. In addition, modeling the pose estimation pipeline as a random effect captures algorithmic variability in the measurement process. Consequently, \eqref{meta_regress} partitions the total variance into within-participant, between-participant, between-exercise, and between-pipeline components, thereby accounting for the repeated-measures structure of the original experimental design and the variability introduced by using multiple deep learning architectures. The inclusion of inverse-variance weights further ensures that observations with more precise performance estimates (e.g., derived from a larger number of repetitions or lower measurement error) exert greater influence on the estimation of the fixed and random effects.

To assess confidence in the chosen empirical specification, we compared several alternatives to Model~\eqref{meta_regress}. In particular, we evaluated models that (i) included only random intercepts; and (ii) employed different covariance structures, including specifications allowing correlations between intercept and slope effects \citep{viechtbauer2015package}. Model~\eqref{meta_regress} was retained because it provided the best balance of fit and parsimony, achieving the lowest AIC and BIC values while avoiding poorly identified covariance parameters \citep{pinheiro2000mixed}. Importantly, fixed-effect estimates were qualitatively consistent across specifications, including models with different covariance structures and those omitting the pose estimation pipeline as a random effect.

\section{Results and Post-Hoc Analyses} \label{sec:results}
In Section~\ref{sec:accuracy}, we assess the accuracy of the pose estimation pipeline. We then present three analyses based on Model \eqref{meta_regress}. Section~\ref{section1} examines kinematic contrasts between ROM conditions to characterize features of lengthened partials beyond mechanical reductions in ROM. Section~\ref{section2} evaluates the effect of pROM on between-participant and between-exercise variability. Finally, Section~\ref{section4} investigates whether a standardized definition of pROM can be established by modeling ROM on a logarithmic scale and estimating a data-driven, exercise-agnostic ROM range.

\begin{table}[t]
\centering
\scalebox{0.9}{\begin{tabular}{>{\raggedright\arraybackslash}m{5.5cm}|*{5}{>{\centering\arraybackslash}m{2.2cm}}}
\hline \\
\textbf{Metric}
& \textit{\textbf{RTMPose}}
& \textit{\textbf{RTMO}}
& \textit{\textbf{Landmarker}}
& \textit{\textbf{DWPose}}
& \textit{\textbf{YOLO26}} \\ \\
\hline \\
Number of Videos Included  & 107 & 104 & 110 & 112 & 113 \\ \\
Average Rep Discrepancy & 1.34 & 1.42 & 1.35 &  1.28 & 1.37 \\ \\
Proportion of Videos with Discrepancy $\leq$ 2 & 93.46\% & 93.26\% & 95.45\% & 92.86\% & 91.15\% \\ \\
Proportion of Videos with Discrepancy $\leq$ 1 & 57.00\% & 53.85\% & 58.18\% & 60.71\% & 59.98\% \\ \\
Proportion of Videos with Discrepancy $=$ 0 & 17.76\% & 10.57\% &10.91\% & 20.54\%  &  17.70\% \\ \\
Max Repetition Discrepancy & 4 & 3 & 3 & 4 & 3 \\ \\  \hline \\
Number of Videos Included & 258 & 252 & 274 & 272 & 276 \\ \\
Mean ROM Reduction & 3.60\% & 37.34\% & 3.77\% & 14.80\% & 9.52\%  \\ \\
Mean Duration Reduction & 0.63s & 0.98s & 1.23s & 1.33s & 0.57s \\ \\
\hline
\end{tabular}}
\caption{Performance metrics for each pose estimation method. The top represents results on the validation set (119 annotated videos); the final three rows report results for the full dataset (303 videos).}
\label{tab:validation_metrics}
\end{table}

\subsection{Pipeline Accuracy and ROM Detection} \label{sec:accuracy}

We first note that establishing a ground truth for range of motion and repetition counts is conceptually non-trivial. The primary challenge resides not in the visual detection of movement, but in the definitional ambiguity of what constitutes an \textit{intentional} repetition within a specific ROM condition. For instance, determining the precise boundaries of a set requires distinguishing valid repetitions from ancillary movements, such as grip adjustments, postural shifts, or incidental partial repetitions. Consequently, manual annotation relies on interpretive judgment, illustrating that validating pipeline performance is fundamentally constrained by the conceptual framework used to define the exercise construct rather than a purely mechanical estimation of displacement.

Notwithstanding the subjectivity inherent in manual observation, the dataset was standardized by labeling all 303 non-duplicated videos according to exercise type, ROM condition, and exercising side. Of these, 119 videos (40\%) from 11 participants were manually annotated to verify the number of exercise repetitions and used as a validation set. Hyperparameters were optimized using Optuna through 3,000 cross-validated iterations. Applying the selected parameters to the validation set, the pose estimation pipelines (Table~\ref{tab:validation_metrics}) achieved an average absolute repetition discrepancy between 1.28 and 1.42 repetitions relative to the ground truth while retaining between 252 and 276 videos from the full dataset. Agreement within $\pm2$ repetitions ranged from 91-95\% across models, and no discrepancy exceeded four repetitions. Thus, repetition detection was consistent across models.

While repetition detection remained consistent across models, the resulting kinematic summaries on the full dataset do exhibit variability, particularly in estimates of average ROM reduction and repetition-duration differences between pROM and fROM (bottom of Table~\ref{tab:validation_metrics}). Such variation is expected given differences in model architectures and keypoint detection strategies, which can influence joint-angle reconstruction and temporal smoothing. Nevertheless, despite these differences, the models produced comparable kinematic summaries and preserved the same qualitative contrasts between pROM and fROM conditions. The consistency of these patterns across pose estimation architectures suggests that the downstream findings are not driven by the idiosyncrasies of any single model. These observations align with prior work showing that markerless pose estimation systems can recover joint kinematics with high accuracy \citep[e.g.,][]{balci2025reliability, scataglini2025systematic, carriere2025can}. Thus, the results indicate that the proposed pipeline provides a reliable and architecture-agnostic approach for extracting resistance training kinematics.

\begin{table}[b]
\centering
\begin{tabular}{l|cc|cc}
\hline
\multirow{2}{*}{ \hspace*{\fill} \textbf{Exercise}} & \multicolumn{2}{c|}{\textbf{Full ROM}} & \multicolumn{2}{c}{\textbf{Partial ROM}} \\
 & \textbf{Videos} & \textbf{Repetitions} & \textbf{Videos} & \textbf{Repetitions} \\
\hline
Bayesian Curl & 16 & 114 & 18 & 155 \\
Cable Pushdown & 22 & 160 & 21 & 154 \\
Dumbbell Curl & 16 & 146 & 17 & 172 \\
Overhead Extension & 14 & 140 & 14 & 174 \\
Dumbbell Row  & 17 & 157 & 17 & 162 \\
Flat Press  & 16 & 111 & 17 & 155 \\
Incline Press  & 17 & 167 & 16 & 160 \\
Lat Pulldown & 21 & 144 & 18 & 145 \\ \hline
Total  & 135 & 1106 & 137 & 1290 \\
\hline
\end{tabular}
\caption{Number of analyzed videos/repetitions per exercise and ROM condition (\textit{DWPose}).}
\label{tab:rom_counts}
\end{table}

As \textit{DWPose} produced repetition-detection accuracy and downstream kinematic estimates that were somewhat representative of the overall ensemble while retaining a large proportion of usable videos (Table~\ref{tab:validation_metrics}), we used for the reporting of descriptive statistics. Table~\ref{tab:rom_counts} summarizes the final data sample under this specification, comprising 272 analyzed videos and 2,396 detected repetitions across eight exercises and two ROM conditions. The distribution of observations is broadly balanced, although the number of repetitions varies due to exercise-specific programming and adjustments to training intensity. Table~\ref{table:summary} reports the corresponding descriptive statistics stratified by ROM condition. As expected, average ROM is consistently greater under fROM than pROM across all exercises, with particularly pronounced reductions for pressing and pushdown movements. Repetition durations are broadly comparable between fROM and pROM, with no consistent directional pattern, while eccentric and concentric phase durations vary across exercises. The number of repetitions per set also varies across exercises and ROM conditions, consistent with the heterogeneous repetition ranges prescribed in the training protocol. Finally, body-side proportions are broadly balanced, indicating no systematic bias toward left- or right-sided execution.

\begin{table}[t]
\footnotesize
\centering
\begin{tabular}{lcccccc}
\toprule
 & \textbf{Average} & \textbf{Rep} & \textbf{Eccentric} & \textbf{Concentric} & \textbf{Exercise} & \textbf{Body} \\
\textbf{Exercise} & \textbf{ROM (°)} &  \textbf{Duration (s)} & \textbf{Duration (s)} &  \textbf{Duration (s)} & \textbf{Repetitions} & \textbf{Side} \\
\midrule
Bayesian Curl & 58.2 (42.6) & 3.5 (1.1) & 1.9 (0.7) & 1.6 (0.5) & 8.7 (3.5) & 0.62 \\
Cable Pushdown & 79.1 (47.5) & 3.8 (1.1) & 1.6 (0.6) & 2.2 (0.7) & 7.6 (1.6) & 0.67 \\
Dumbbell Curl & 79.2 (54.1) & 4.3 (0.8) & 2.4 (0.6) & 1.9 (0.5) & 10.4 (3.0) & 0.38 \\
Overhead Extension & 129.9 (36.9) & 4.1 (0.8) & 1.7 (0.4) & 2.3 (0.6) & 11.8 (4.5) & 0.43 \\
Dumbbell Row & 89.8 (20.0) & 3.6 (0.5) & 2.2 (0.5) & 1.4 (0.3) & 10.1 (2.1) & 0.41 \\
Flat Press & 83.8 (26.5) & 3.9 (0.6) & 1.8 (0.4) & 2.1 (0.5) & 8.1 (2.8) & 0.38 \\
Incline Press & 111.1 (47.7) & 3.2 (0.7) & 1.3 (0.7) & 1.9 (0.7) & 10.8 (2.5) & 0.47 \\
Lat Pulldown & 128.9 (26.1) & 3.9 (0.7) & 2.5 (0.6) & 1.4 (0.3) & 7.4 (1.6) & 0.62 \\
\midrule
Bayesian Curl & 45.0 (29.8) & 3.6 (0.8) & 1.5 (0.7) & 2.1 (0.7) & 10.3 (3.8) & 0.44 \\
Cable Pushdown & 30.9 (25.3) & 3.1 (0.6) & 1.3 (0.4) & 1.7 (0.6) & 10.7 (4.9) & 0.45 \\
Dumbbell Curl & 39.8 (35.1) & 3.8 (0.9) & 1.9 (0.7) & 1.9 (0.5) & 10.7 (2.1) & 0.59 \\
Overhead Extension & 83.7 (33.1) & 3.5 (0.6) & 1.5 (0.3) & 2.0 (0.4) & 15.4 (6.5) & 0.64 \\
Dumbbell Row & 75.6 (11.5) & 3.4 (0.4) & 2.1 (0.2) & 1.3 (0.3) & 11.3 (4.6) & 0.65 \\
Flat Press & 43.7 (18.8) & 3.3 (0.6) & 1.9 (0.6) & 1.4 (0.5) & 11.2 (4.1) & 0.65 \\
Incline Press & 34.0 (30.3) & 2.9 (0.4) & 1.3 (0.4) & 1.6 (0.5) & 10.6 (2.2) & 0.50 \\
Lat Pulldown & 91.4 (21.3) & 3.8 (0.6) & 2.3 (0.4) & 1.5 (0.4) & 9.4 (3.1) & 0.39 \\
\bottomrule
\end{tabular}
\caption{Descriptive statistics (DWPose): fROM - top; pROM - bottom. Values are weighted means (standard deviations) based on repetition numbers. Body Side is the proportion of exercises performed using the left arm.}
\label{table:summary}
\end{table}

\subsection{Execution Patterns and Temporal Dynamics} \label{section1}
Let $Y_{iegm}$ denote the repetition or phase duration of each exercise and ROM condition. By estimating Model \eqref{meta_regress}, the sign and magnitude of $\beta_1$ indicates whether pROM is associated with systematic changes in these metrics. A significant effect would suggest that performing lengthened partials influences not only ROM (by definition), but also the overall execution of the different exercises.

\begin{table}[ht]
\centering
\small
\setlength{\tabcolsep}{6pt}
\begin{tabular}{l cc cc cc cc}
\toprule
  & \multicolumn{2}{c}{Rep Duration (s)} & \multicolumn{2}{c}{Eccentric Phase (s)} & \multicolumn{2}{c}{Concentric Phase (s)} & \multicolumn{2}{c}{Range of Motion ($^\circ$)} \\
\cmidrule(lr){2-3} \cmidrule(lr){4-5} \cmidrule(lr){6-7} \cmidrule(lr){8-9}
  & Estimate & SE & Estimate & SE & Estimate & SE & Estimate & SE \\
\midrule
\addlinespace
\multicolumn{9}{l}{\textbf{Fixed Effects}} \\
Intercept & 3.569$^{***}$ & 0.249 & 1.880$^{***}$ & 0.182 & 1.603$^{***}$ & 0.132 & 89.851$^{***}$ & 10.154 \\
pROM & -0.035 & 0.281 & -0.137 & 0.236 & -0.002 & 0.175 & -39.954$^{***}$ & 9.310 \\
Female & 0.141 & 0.239 & 0.059 & 0.106 & 0.147 & 0.114 & -2.217 & 5.133 \\
\addlinespace
\multicolumn{9}{l}{\textbf{Random Effects}} \\
$\tau_{p}^2$ & 0.643 &  & 0.067 &  & 0.093 &  & 455.064 &  \\
$\tau_{q}^2$ & 0.275 &  & 0.087 &  & 0.082 &  & 110.741 &  \\
$\tau_{u}^2$ & 0.202 &  & 0.239 &  & 0.102 &  & 342.401 &  \\
$\tau_{v}^2$ & 0.146 &  & 0.160 &  & 0.089 &  & 176.896 &  \\
$\tau_{r}^2$ & 0.052 &  & 0.002 &  & 0.004 &  & 210.822 &  \\
\addlinespace
\multicolumn{9}{l}{\textbf{Model Fit}} \\
$Q_E$ & 138841.20$^{***}$ &  & 121666.21$^{***}$ &  & 68138.38$^{***}$ &  & 1414905.24$^{***}$ &  \\
$Q_M$ & 0.18 &  & 0.32 &  & 0.83 &  & 9.30$^{***}$ &  \\
\bottomrule
\end{tabular}
\caption{Results associated with estimating a crossed random-effects regression with participant and exercise random intercepts/slopes and a pose estimation-pipeline random intercept. The within-cell sampling variance was specified as $s^2/k$. Significance codes: `***': $<0.001$, `**': $<0.01$, `*': $<0.05$, `.' : $<0.1$.}
\label{table:exercise_executionDIAG}
\end{table}

Table~\ref{table:exercise_executionDIAG} reports the crossed random-effects regression results for all four outcomes. After adjusting for sex, the pROM condition was associated with a large and statistically significant reduction in range of motion ($\beta_1=-39.95^\circ$, SE $=9.31$, $p<0.001$). In contrast, the estimated effects of pROM on repetition duration ($\beta_1=-0.035$, SE $=0.281$), eccentric phase duration ($\beta_1=-0.137$, SE $=0.236$), and concentric phase duration ($\beta_1=-0.002$, SE $=0.175$) were small and not statistically significant. The sex coefficient was also small and nonsignificant across all outcomes.

The random-effects estimates indicate heterogeneity across both participants and exercises. For repetition duration, participant-level variability exceeded exercise-level variability for both intercepts ($\tau_p^2=0.643$ vs.\ $\tau_u^2=0.202$) and slopes ($\tau_q^2=0.275$ vs.\ $\tau_v^2=0.146$), suggesting that differences in overall repetition tempo and its response to pROM were more strongly associated with individuals than with the exercise performed. For the phase-specific outcomes, exercise-level variability was more pronounced. That is, in the eccentric phase model, exercise-level intercept and slope variances ($\tau_u^2=0.239$, $\tau_v^2=0.160$) exceeded the corresponding participant-level variances ($\tau_p^2=0.067$, $\tau_q^2=0.087$), with a similar pattern for the concentric phase. For range of motion, substantial variability was observed at both the participant ($\tau_p^2=455.064$, $\tau_q^2=110.741$) and exercise levels ($\tau_u^2=342.401$, $\tau_v^2=176.896$), indicating that both individuals and exercises contributed meaningfully to differences in baseline ROM and the magnitude of ROM reductions under pROM. Finally, variance attributable to the pose estimation pipelines, although larger for ROM, were small relative to participant- and exercise-level variability for all outcomes, suggesting that most variability reflected differences in movement execution rather than measurement errors across architectures.

Moderator tests ($Q_M$) indicated that the included predictors explained a statistically significant portion of variability only for the ROM outcome ($Q_M=9.30$, $p<0.001$), whereas the models for repetition duration and concentric/eccentric phase durations did not show significant moderator effects. Substantial residual heterogeneity remained across all outcomes ($Q_E$), suggesting that there were additional unmeasured factors that contributed to variation in movement execution.

Overall, the results confirm that pROM is associated with a marked reduction in ROM while (somewhat counterintuitively) no decrease in repetition duration. Because participants in \citet{wolf2025lengthened} were prescribed load adjustments to maintain target repetition ranges across ROM conditions, this pattern is consistent with the intended experimental constraint. Nevertheless, more broadly, these findings suggest that the comparable hypertrophy and strength adaptations reported in prior work \citep{moreno2024does, varovic2025effects,haverspartial} may partly reflect similar per-repetition time under tension despite the reductions in ROM \citep[e.g.,][]{burd2012muscle}.

\subsection{Variance Decomposition Analysis} \label{section2}
We next examined whether variability differed between ROM conditions. Since the pose estimation pipeline enters as a random intercept only, it contributes equally to variability under both pROM and fROM. Consequently, the subsequent ROM-specific variance comparison focused on comparing participant- and exercise-level random effects, which include condition-specific random slope terms. 

Based on Model~\eqref{meta_regress}, the participant-level variability under each ROM condition is given by
\begin{equation*}
V_{p,\mathrm{fROM}} = \operatorname{Var}(p_i) = \tau_p^2, \qquad
V_{p,\mathrm{pROM}} = \operatorname{Var}(p_i + q_i)
= \tau_p^2 + \tau_q^2.
\end{equation*}
where the equality follows by independence. Similarly, the exercise-level variability is 
\begin{equation*}
V_{e,\mathrm{fROM}} = \operatorname{Var}(u_e) = \tau_u^2,
\qquad
V_{e,\mathrm{pROM}} = \operatorname{Var}(u_e + v_e)
= \tau_u^2 + \tau_v^2.
\end{equation*}
As a consequence, we defined the variance contrasts
\begin{equation*}
D_p
=
V_{p,\mathrm{pROM}} - V_{p,\mathrm{fROM}}
=
\tau_q^2,
\qquad
D_e
=
V_{e,\mathrm{pROM}} - V_{e,\mathrm{fROM}}
=
\tau_v^2,
\end{equation*}
which quantify the change in between-participant and between-exercise variability, respectively, when performing lengthened partials versus full range of motion. One-sided hypothesis tests were then used to evaluate whether variability under pROM exceeded that of fROM, i.e., 
\begin{equation*}
H_0^{(p,+)}: D_p \le 0
\quad \text{vs.} \quad
H_A^{(p,+)}: D_p > 0,
\qquad 
H_0^{(e,+)}: D_e \le 0
\quad \text{vs.} \quad
H_A^{(e,+)}: D_e > 0.
\end{equation*}

Inference was conducted using likelihood ratio tests (LRT) for tractability, comparing \eqref{meta_regress} to nested models in which the corresponding random-slope variance was constrained to be zero (i.e., $\tau_q^2 = 0$ for participants and $\tau_v^2 = 0$ for exercises). Because variance components are non-negative, the null hypotheses lie on the boundary of the parameter space. Thus, we used the boundary-adjusted asymptotic distribution, computing one-sided p-values as $\tfrac{1}{2}P(\chi^2_1 \ge \text{LRT})$. 
A statistically significant result indicated increased variability under pROM for the corresponding level.

\begin{table}[ht]
\centering
\begin{tabular}{lcccc}
\toprule
\textbf{Contrast} & \textbf{Statistic} & \textbf{Estimate} & \textbf{LRT} & \textbf{$p$-value (one-sided)} \\
\midrule
\multicolumn{5}{c}{\textbf{Repetition Duration (s)}}
\\
Participant-Level Variance & $D_p$ & 0.275 & 6317.514 & $p<0.001^{***}$  \\
Exercise-level Variance    & $D_e$ & 0.146 & 3094.253 & $p<0.001^{***}$ \\
\midrule
\multicolumn{5}{c}{\textbf{Range of Motion (°)}}
\\
Participant-Level Variance & $D_p$  & 110.810 & 44352.794 & $p<0.001^{***}$ \\
Exercise-level Variance    & $D_e$  & 176.605 & 38053.138 & $p<0.001^{***}$\\
\bottomrule
\end{tabular}
\caption{Likelihood ratio tests for differences in variation. One-sided $p$-values test whether the indicated variance contrast is less than zero. Significance codes: ‘***’:$<0.001$, ‘**’:$<0.01$, ‘*’:$<0.05$, ‘.’ : $<0.1$.}
\label{table:contrasts}
\end{table}

The results are presented in Table~\ref{table:contrasts} for repetition duration (s) and ROM (°). For repetition duration, both the participant-level ($D_p = 0.275$) and exercise-level ($D_e = 0.146$) variance contrasts were positive and highly statistically significant ($p<0.001$), indicating greater between-participant and between-exercise variability under pROM relative to fROM. Similarly, for ROM, both contrasts were positive and statistically significant ($D_p = 110.810$, $D_e = 176.605$, both $p<0.001$), again indicating increased variability under pROM. These results provide strong evidence that the corresponding variance component is positive, implying that lengthened partials are associated with greater variability in both repetition duration and ROM at both the participant and exercise levels.

\subsection{Towards the Standardization of Lengthened Partial Movements} \label{section4}
We next introduce \%ROM, defined as the percentage of fROM achieved during a lengthened partial repetition. We compared this metric across exercises to assess whether lengthened partials admit an absolute definition or vary with each exercise’s strength curve \citep[e.g.,][]{newmire2018partial}. This analysis informs whether an \textit{ideal} ROM proportion exists for lengthened partial training.

For participant $i$, let $R_{ieg}$ denote the range of motion across repetitions for exercise $e$ and ROM condition $g$. To compute \%ROM, we analyzed the log-transformed outcomes
\begin{equation*}
Y_{ieg} = \log(R_{ieg}),
\end{equation*}
which allowed proportional differences between pROM and fROM to be modeled on a multiplicative scale. Under this transformation, mean ROM corresponded to a weighted geometric proportion rather than the weighted arithmetic mean reported in Table~\ref{table:summary}. However, the coefficient $\beta_1$ from Model~\eqref{meta_regress} now represents the log ratio of pROM to fROM. Sampling variances were defined via the delta method as $\sigma_{ieg}^2 = s^2_{ieg}/(k_{ieg} R_{ieg}^2)$ and were used to construct inverse-variance weights.

As compared with Table~\ref{table:exercise_executionDIAG}, the pROM effect remained statistically significant ($\beta_1=-0.5733$, $SE=0.1311$, $p<0.001$), indicating a substantial reduction in range of motion under pROM. The moderator test was also significant ($Q_M=9.8939$, $p<0.001$), while residual heterogeneity remained large ($Q_E=772991.521$, $p<0.001$), suggesting additional unexplained variation. To assess whether the pROM condition had exercise-specific variation, we examined the variance of the random slope, $\tau_v^2$. A likelihood ratio test comparing Model~\eqref{meta_regress} with a reduced model without exercise-specific slopes indicated that this variance component was statistically significant (LRT$=28635.85$, $p<0.001$), providing initial evidence that the effect of pROM differed across exercises.

To further characterize these differences, we defined
\begin{equation} \label{exercise_level}
\delta_e = \beta_1 + v_e,
\end{equation}
where $v_e$ denotes the exercise-specific random slope associated with the pROM indicator. Accordingly, for exercise $e$, $\delta_e$ represents the log ratio of pROM to fROM, combining the overall mean effect $\beta_1$ with an exercise-specific deviation $v_e$. Thus, $\%ROM_e = 100\exp(\delta_e)$ provides the corresponding percentage of fROM achieved under pROM. Exercise-specific effects were estimated using best linear unbiased predictions (BLUPs), which shrank exercise-level estimates toward the overall mean, thereby stabilizing estimates in the presence of unequal sample sizes. Confidence intervals were obtained using Wald approximations based on the estimated standard errors of $\delta_e$.

\begin{table}[ht]
\centering
\small
\begin{tabular}{lccc}
\toprule
\textbf{Exercise} & \textbf{$\%$ROM }& \textbf{$95\%$ CI} & \textbf{$p$ (Wald)}\\
\midrule
Bayesian Curl & 58.5 & {}[42.3, 80.8] & 0.717\\
Cable Pushdown & 54.7 & {}[39.6, 75.6] & 0.762\\
Dumbbell Curl & 63.4 & {}[45.8, 87.6] & 0.244\\
Overhead Extension & 82.3 & {}[59.5, 113.8] & $<$0.001\\
Dumbbell Row & 62.4 & {}[45.2, 86.3] & 0.310\\
Flat Press & 43.3 & {}[31.3, 59.9] & 0.009\\
Incline Press & 32.4 & {}[23.4, 44.8] & $<$0.001\\
Lat Pulldown & 69.8 & {}[50.5, 96.4] & 0.034\\ 
\bottomrule
\end{tabular}
\caption{Exercise-specific estimates of pROM relative to fROM using Wald confidence intervals and best linear unbiased predictions (BLUPs). We report the effect size ($\%$ROM), 95\% confidence interval of the effect size, and the outcome ($p$-value) of a two-sided test to determine whether the indicated contrast $\delta_e$ differs from $\beta_1$. \\ Significance codes: ‘***’:$<0.001$, ‘**’:$<0.01$, ‘*’:$<0.05$, ‘.’ : $<0.1$.}
\label{table:deviations}
\end{table}

To summarize the expected variation across exercises, we constructed a model-based prediction interval for the exercise-specific effects, $\delta_e \sim \mathcal{N}(\beta_1, \tau_v^2)$. On the percentage scale, this corresponded to a 95\% prediction interval of $[40\%,80\%]$, centered around a mean of $56\%$. This interval captured the range of \%ROM values expected across exercises under the fitted model. While several exercises fell within this range, notable deviations occurred at both extremes, particularly for pressing movements (lower \%ROM) and certain pulling or extension movements (higher \%ROM). In particular, the exercise-specific effects of pROM reported in Table~\ref{table:deviations} exhibit substantial heterogeneity. The Incline Press and Flat Press exhibited markedly lower \%ROM than the overall mean, with estimates of $32.4\%$ ($p<0.001$) and $43.3\%$ ($p=0.009$), respectively. In contrast, the Overhead Extension demonstrated a substantially higher \%ROM ($82.3\%$, $p<0.001$), while the Lat Pulldown also showed a higher-than-average value ($69.8\%$, $p=0.034$). The remaining exercises (i.e., Bayesian Curl, Cable Pushdown, Dumbbell Curl, and Dumbbell Row) were not statistically distinguishable from the overall mean, but had mean estimates ranging from $54.7\%$ to $63.4\%$.

Finally, to contextualize these findings, we conducted pairwise comparisons of the exercise-specific effects. Specifically, using the BLUP-based estimates from \eqref{exercise_level}, Wald-type tests were used to evaluate the null hypothesis that $H_0:\,\delta_e = \delta_{e'}$ for all $e \neq e'$, with $p$-values adjusted using the Holm–Bonferroni procedure \citep{abdi2010holm}. Exercises were grouped into clusters such that all within-cluster pairwise differences were not statistically significant after correction. This analysis identified three clusters: (i) Bayesian Curl, Cable Pushdown, Dumbbell Curl, and Dumbbell Row (mean $=59.7\%$, range: $54.7$–$63.4\%$); (ii) Overhead Extension and Lat Pulldown (mean $=76.0\%$, range: $69.8$–$82.3\%$); and (iii) Flat Press and Incline Press (mean $=37.9\%$, range: $32.4$–$43.3\%$). The clustering approach suggested that the exercise-specific effects formed multiple distinct groupings.

Practically, the results in this section strongly indicate that the proportion of full ROM achieved under pROM may not be invariant across exercises. Although the average effect suggests that lengthened partials correspond to approximately $56\%$ of full ROM, a wide prediction interval and substantial between-exercise variability suggests that no single, precise, exercise-agnostic definition exists. Instead, lengthened partials may be better characterized as an exercise-dependent construct, with systematic variation reflecting underlying biomechanical differences across movements.

\section{Discussion and Conclusion} \label{sec:conclusion}
This paper summarizes the development of an AI-based pose estimation pipeline to analyze video data from \citet{wolf2025lengthened}, comprising eight upper-body resistance exercises performed by 26 participants under both lengthened partial (pROM) and full range of motion (fROM) conditions. Leveraging the crossed design of the original study, we compared range of motion (ROM) and repetition duration between pROM and fROM. We found that pROM reduces ROM without shortening duration and that both between-participant and between-exercise variability increased under lengthened partials relative to full ROM. To facilitate cross-exercise comparisons, we introduced \%ROM, defined as the proportion of fROM achieved during pROM. Using model-based estimates that account for participant- and exercise-level heterogeneity, we found that pROM corresponded to approximately $56\%$ of full ROM but varied systematically across exercises, suggesting that for upper-body movements, lengthened partials may not admit a consistent empirical definition.

The use of AI-based pose estimation has gained prominence in domains such as fall detection for older adults \citep{chang2021pose}, exercise recognition for physiotherapy \citep{raza2023logrf}, 
multi-agent motion capture in professional sports settings such as basketball \citep{williams2026effects}, and workout tracking in AI-driven fitness applications \citep{kaushik2024ai}. In this study, it is applied to address unresolved questions in the science-based strength training community regarding the definition and characterization of a particular training modality \citep{milo_Wolf_2023, schofield_2024, eugeneteo_2024, meno_2025}. By leveraging the high-resolution data derived from the AI pipelines, we analyzed key aspects of exercise execution, the propagation of variability across ROM conditions, and the extent to which a common empirical definition of lengthened partials could be established across exercises. Ultimately, this methodology represents a unique application of computer vision, providing a novel and accurate means of quantifying and better understanding the granular execution differences that define modern resistance training protocols \citep{scataglini2024accuracy}.

The statistical findings derived from the pose estimation pipeline have important implications for prescribing and communicating lengthened partials in resistance training. For instance, although by definition, pROM involves a reduced range of motion, our results suggest that load adjustments to maintain a similar per-repetition time under tension as fROM may also be important for achieving comparable hypertrophic adaptations \citep[e.g.,][]{varovic2025effects, haverspartial}. Furthermore, the greater between-participant and between-exercise variability observed under pROM highlights the need for more precise coaching cues to ensure consistent execution. That is, unlike fROM, which is typically bounded by clear anatomical or mechanical end-points that provide proprioceptive feedback \citep{riemann2002sensorimotor}, pROM requires the trainee to reverse the movement mid-range, increasing susceptibility to inter-repetition variation \citep[e.g.,][]{pallares2021effects}. As a consequence, the findings in this research emphasize the importance of rigorous technique monitoring and individualized load adjustments to ensure consistency in partial-range training.

The AI pipeline, while objective and reproducible, has several limitations. First, it relies on monocular 2D/3D pose estimation, which infers body landmarks using deep learning models trained on large, heterogeneous datasets. Although robust to many real-world imaging conditions, these models remain sensitive to occlusion, truncation, and
inconsistent framing \citep{colyer2018review}. This was evident in our pipeline, as the video dataset was not originally collected with AI-based motion analysis in mind. As a result, many videos in which repetitions were clearly identifiable to human observers had to be excluded from further analysis. This highlights a broader methodological challenge of retrospective video analysis. Future studies should adopt recording protocols optimized for pose estimation, such as standardized camera placement, clear visibility, and controlled lighting.

Second, joint angles could not be anatomically validated. In the absence of synchronized marker-based motion capture or other ground-truth kinematic measurements, it is not possible to directly assess joint-angle accuracy or identify systematic biases. Accordingly, the reported ROM values should be interpreted as kinematic proxies rather than precise biomechanical measurements. The statistical model accounts for potential measurement variability through a pose estimation random effect, and our analyses primarily rely on within-participant and within-exercise comparisons, which are less sensitive to absolute angle error than to relative consistency. However, future work could strengthen the biomechanical interpretation of these results through laboratory-based validation.

Finally, the dataset is drawn from a single prior study with a limited participant pool, which restricts generalizability. Although the large number of repetitions improves within-subject reliability and statistical power, it does not offset the limited demographic diversity or the relatively narrow range of exercises (e.g., the original study did not include lower-body movements). Given evidence that training prescriptions may differ between upper- and lower-body hypertrophy \citep[e.g.,][]{jung2023muscle}, future work should assess whether the observed variability patterns and the absence of a standardized definition of lengthened partials generalizes across movement types.


While these limitations warrant cautious interpretation, the findings demonstrate that AI-based pose estimation combined with statistical modeling can generate actionable insights that translate the documented hypertrophic benefits of lengthened partial training \citep[e.g.,][]{kassiano2023greater} into practical guidance. The results identify key execution characteristics of lengthened partials that can inform practitioners, while the greater observed variability underscores the need for precise technique monitoring, individualized coaching, and, where appropriate, exercise-specific cues or loading strategies. More broadly, this study highlights the potential of AI-driven kinematic analysis to bridge the gap between controlled experimental findings and real-world training practice.


\bibliographystyle{informs2014} 
\bibliography{references}

@article{wolf2025lengthened,
  title={Lengthened partial repetitions elicit similar muscular adaptations as full range of motion repetitions during resistance training in trained individuals},
  author={Wolf, Milo and Korakakis, Patroklos Androulakis and Pi{\~n}ero, Alec and Mohan, Adam E and Hermann, Tom and Augustin, Francesca and Sapuppo, Max and Lin, Brian and Coleman, Max and Burke, Ryan and others},
  journal={PeerJ},
  volume={13},
  pages={e18904},
  year={2025},
  publisher={PeerJ Inc.}
}

@inproceedings{lugaresi2019mediapipe,
  title={Mediapipe: A framework for perceiving and processing reality},
  author={Lugaresi, Camillo and Tang, Jiuqiang and Nash, Hadon and McClanahan, Chris and Uboweja, Esha and Hays, Michael and Zhang, Fan and Chang, Chuo-Ling and Yong, Ming and Lee, Juhyun and others},
  booktitle={Third workshop on computer vision for AR/VR at IEEE computer vision and pattern recognition (CVPR)},
  volume={2019},
  year={2019}
}

@article{schafer2011savitzky,
  author={Schafer, Ronald W.},
  journal={IEEE Signal Processing Magazine}, 
  title={What Is a Savitzky-Golay Filter?}, 
  year={2011},
  volume={28},
  number={4},
  pages={111-117}
}

@article{maeo2023triceps,
  title={Triceps brachii hypertrophy is substantially greater after elbow extension training performed in the overhead versus neutral arm position},
  author={Maeo, Sumiaki and Wu, Yuhang and Huang, Meng and Sakurai, Hikaru and Kusagawa, Yuki and Sugiyama, Takashi and Kanehisa, Hiroaki and Isaka, Tadao},
  journal={European journal of sport science},
  volume={23},
  number={7},
  pages={1240--1250},
  year={2023},
  publisher={Taylor \& Francis}
}

@article{pedrosa2023training,
  title={Training in the initial range of motion promotes greater muscle adaptations than at final in the arm curl},
  author={Pedrosa, Gustavo F and Sim{\~o}es, Marina G and Figueiredo, Marina OC and Lacerda, Lucas T and Schoenfeld, Brad J and Lima, Fernando V and Chagas, Mauro H and Diniz, Rodrigo CR},
  journal={Sports},
  volume={11},
  number={2},
  pages={39},
  year={2023},
  publisher={MDPI}
}

@article{pedrosa2022partial,
  title={Partial range of motion training elicits favorable improvements in muscular adaptations when carried out at long muscle lengths},
  author={Pedrosa, Gustavo F and Lima, Fernando V and Schoenfeld, Brad J and Lacerda, Lucas T and Sim{\~o}es, Marina G and Pereira, Mariano R and Diniz, Rodrigo CR and Chagas, Mauro H},
  journal={European journal of sport science},
  volume={22},
  number={8},
  pages={1250--1260},
  year={2022},
  publisher={Taylor \& Francis}
}

@article{kassiano2023greater,
  title={Greater gastrocnemius muscle hypertrophy after partial range of motion training performed at long muscle lengths},
  author={Kassiano, Witalo and Costa, Bruna and Kunevaliki, Gabriel and Soares, Danrlei and Zacarias, Gabriel and Manske, Ingrid and Takaki, Yudi and Ruggiero, Maria Fernanda and Stavinski, Nat{\~a} and Francsuel, Jarlisson and others},
  journal={The Journal of Strength \& Conditioning Research},
  volume={37},
  number={9},
  pages={1746--1753},
  year={2023},
  publisher={LWW}
}

@article{androulakis2024optimizing,
  title={Optimizing resistance training technique to maximize muscle hypertrophy: A narrative review},
  author={Androulakis Korakakis, Patroklos and Wolf, Milo and Coleman, Max and Burke, Ryan and Pi{\~n}ero, Alec and Nippard, Jeff and Schoenfeld, Brad J},
  journal={Journal of Functional Morphology and Kinesiology},
  volume={9},
  number={1},
  pages={9},
  year={2024},
  publisher={Multidisciplinary Digital Publishing Institute}
}

@article{ottinger2023muscle,
  title={Muscle hypertrophy response to range of motion in strength training: a novel approach to understanding the findings},
  author={Ottinger, Charlie R and Sharp, Matthew H and Stefan, Matthew W and Gheith, Raad H and de la Espriella, Fernando and Wilson, Jacob M},
  journal={Strength \& Conditioning Journal},
  volume={45},
  number={2},
  pages={162--176},
  year={2023},
  publisher={LWW}
}

@article{kassiano2023roms,
  title={Which ROMs lead to Rome? A systematic review of the effects of range of motion on muscle hypertrophy},
  author={Kassiano, Witalo and Costa, Bruna and Nunes, Joao Pedro and Ribeiro, Alex S and Schoenfeld, Brad J and Cyrino, Edilson S},
  journal={The Journal of Strength \& Conditioning Research},
  volume={37},
  number={5},
  pages={1135--1144},
  year={2023},
  publisher={LWW}
}

@article{newmire2018partial,
  title={Partial compared with full range of motion resistance training for muscle hypertrophy: A brief review and an identification of potential mechanisms},
  author={Newmire, Daniel E and Willoughby, Darryn S},
  journal={The Journal of Strength \& Conditioning Research},
  volume={32},
  number={9},
  pages={2652--2664},
  year={2018},
  publisher={LWW}
}

@article{varovic2025effects,
  title={The Effects of Long Muscle Length Isometric versus Full Range of Motion Isotonic Training on Regional Quadriceps Femoris Hypertrophy in Resistance-Trained Individuals},
  author={Varovic, Dorian and Zganjer, Kristian and Wolf, Milo and Androulakis-Korakakis, Patroklos and Schoenfeld, Brad Jon and Mikulic, Pavle},
  journal={Applied Physiology, Nutrition, and Metabolism},
  number={ja},
  year={2025}
}

@article{viechtbauer2015package,
  title={Package ‘metafor’},
  author={Viechtbauer, Wolfgang and Viechtbauer, Maintainer Wolfgang},
  journal={The comprehensive R Archive network. Package ‘metafor},
  year={2015}
}

@Manual{rprogramming,
    title = {R: A Language and Environment for Statistical Computing},
    author = {{R Core Team}},
    organization = {R Foundation for Statistical Computing},
    address = {Vienna, Austria},
    year = {2024},
    url = {https://www.R-project.org/},
  }

@inproceedings{kaushik2024ai,
  title={AI-Based Posture Correction, Real-Time Exercise Tracking and Feedback using Pose Estimation Technique},
  author={Kaushik, M and Kandala, Mahita and Vignesh, GS and Vithyatharshana, N and Palaniswamy, Suja},
  booktitle={2024 International Conference on Communication, Control, and Intelligent Systems (CCIS)},
  pages={1--6},
  year={2024},
  organization={IEEE}
}

@article{raza2023logrf,
  title={LogRF: An approach to human pose estimation using skeleton landmarks for physiotherapy fitness exercise correction},
  author={Raza, Ali and Qadri, Azam Mehmood and Akhtar, Iqra and Samee, Nagwan Abdel and Alabdulhafith, Maali},
  journal={IEEE Access},
  volume={11},
  pages={107930--107939},
  year={2023},
  publisher={IEEE}
}

@article{chang2021pose,
  title={A pose estimation-based fall detection methodology using artificial intelligence edge computing},
  author={Chang, Wan-Jung and Hsu, Chia-Hao and Chen, Liang-Bi},
  journal={IEEe Access},
  volume={9},
  pages={129965--129976},
  year={2021},
  publisher={IEEE}
}

@misc{milo_Wolf_2023, 
    type={Video}, 
    title={{What are Lengthened Partials? (Science Explained)}}, 
    url={https://www.youtube.com/watch?v=ay59lXp1N_Y}, 
    journal={YouTube}, 
    author={Dr. Milo Wolf}, 
    note = {[Accessed 26-01-2026]},
    year={2023},
}

@misc{eugeneteo_2024, type={Video}, 
    title={{Social Media Fitness is Messed Up - I Need Your Help}}, 
    url={https://www.youtube.com/watch?v=ri0Gr6h2Or4}, 
    journal={YouTube}, 
    author={Eugene Teo}, 
    year={2024}, 
    note = {[Accessed 26-01-2026]},
    language={en} 
}

@misc{meno_2025, type={Video}, 
    title={{6 gorgeous new studies to help you build muscle [2025]}}, url={https://www.youtube.com/watch?v=T2tLFPPdjwY}, 
    journal={YouTube}, 
    author={Menno Henselmans}, 
    year={2025}, 
    note = {[Accessed 26-01-2026]},
    language={en} 
}

@misc{schofield_2024, 
    type={Video}, 
    title={{Lengthened Partials: The Best Way To Get BIG?}}, 
    url={https://www.youtube.com/watch?v=tQJs653JnkE}, 
    journal={YouTube}, 
    author={Geoffrey Schofield}, 
    year={2024}, 
    note = {[Accessed 26-01-2026]},
    language={en} 
}

@article{haverspartial,
  title={Partial Range, Full Gains? The Effect of 8 Weeks of Partial Range of Motion Training at Long Muscle Lengths on Elbow Flexor Hypertrophy and Strength in Trained Individuals},
  author={Havers, Tim and Wagner, Niklas and Held, Steffen and Geisler, Stephan and Wiewelhove, Thimo},
  journal={SportRxiv},
  year={2025}
}

@article{moreno2024does,
  title={Does Performing Resistance Exercise with a Partial Range of Motion at Long Muscle Lengths Maximize Muscle Hypertrophic Adaptations to Training?},
  author={Moreno, Enrique N and Ayers-Creech, Wayne A and Gonzalez, Selena L and Baxter, Holly T and Buckner, Samuel L},
  journal={Journal of Science in Sport and Exercise},
  pages={1--9},
  year={2024},
  publisher={Springer}
}

@book{pinheiro2000mixed,
  title={Mixed-Effects Models in S and S-PLUS},
  author={Pinheiro, José C. and Bates, Douglas M.},
  publisher={Springer},
  address={New York},
  year={2000}
}

@article{viechtbauer2010conducting,
  title={Conducting meta-analyses in {R} with the \texttt{metafor} package},
  author={Viechtbauer, Wolfgang},
  journal={Journal of Statistical Software},
  volume={36},
  number={3},
  pages={1--48},
  year={2010}
}

@article{Nuzzo_2022, title={Narrative review of sex differences in muscle strength, endurance, activation, size, fiber type, and strength training participation rates, preferences, motivations, injuries, and neuromuscular adaptations}, volume={37}, number={2}, journal={Journal of Strength \& Conditioning Research}, author={Nuzzo, James L.}, year={2022}, month={Nov}, pages={494–536}}

@article{mc2021, title={A comparison between male and female athletes in relative strength and power performances}, volume={6}, number={1}, journal={Journal of Functional Morphology and Kinesiology}, author={Bartolomei, Sandro and Grillone, Giuseppe and Di Michele, Rocco and Cortesi, Matteo}, year={2021}, month={Feb}, pages={17}}

@article{wolf2023partial,
  title={Partial vs full range of motion resistance training: A systematic review and meta-analysis},
  author={Wolf, Milo and Androulakis-Korakakis, Patroklos and Fisher, James and Schoenfeld, Brad and Steele, James},
  journal={International Journal of Strength and Conditioning},
  volume={3},
  number={1},
  year={2023}
}

@article{warneke2023physiology,
  title={Physiology of stretch-mediated hypertrophy and strength increases: a narrative review},
  author={Warneke, Konstantin and Lohmann, Lars H and Lima, Camila D and Hollander, Karsten and Konrad, Andreas and Zech, Astrid and Nakamura, Masatoshi and Wirth, Klaus and Keiner, Michael and Behm, David G},
  journal={Sports Medicine},
  volume={53},
  number={11},
  pages={2055--2075},
  year={2023},
  publisher={Springer}
}

@misc{RenaissancePeriodization,
	author = {Israetel, Michael},
	title = {{R}enaissance {P}eriodization},
	howpublished = {\url{https://www.youtube.com/@RenaissancePeriodization}},
	year = {2025},
	note = {[Accessed 22-10-2025]},
}

@misc{youtubeYouTube,
	author = {Nippard, Jeff},
	title = {{Partial Range Of Motion: Broscience Or Legit?}},
	howpublished = {\url{https://www.youtube.com/watch?v=jkaU-mM24_o}},
	year = {2025},
	note = {[Accessed 22-10-2025]},
}

@article{jung2023muscle,
  title={Muscle strength gains per week are higher in the lower-body than the upper-body in resistance training experienced healthy young women—A systematic review with meta-analysis},
  author={Jung, Roger and Gehlert, Sebastian and Geisler, Stephan and Isenmann, Eduard and Eyre, Julia and Zinner, Christoph},
  journal={PloS one},
  volume={18},
  number={4},
  pages={e0284216},
  year={2023},
  publisher={Public Library of Science San Francisco, CA USA}
}

@article{abdi2010holm,
  title={Holm’s sequential Bonferroni procedure},
  author={Abdi, Herv{\'e}},
  journal={Encyclopedia of research design},
  volume={1},
  number={8},
  pages={1--8},
  year={2010},
  publisher={Thousand Oaks, California}
}

@misc{chatgpt_image_2026,
  author       = {{OpenAI}},
  title        = {AI-generated image created using ChatGPT},
  year         = {2026},
  note         = {Image generated by ChatGPT (DALL·E) in response to multipler user prompts},
  howpublished = {\url{https://chat.openai.com/}}
}

@article{orangi2025effects,
  title={The effects of different focus cues and motor learning strategies on landing mechanics in male handball players},
  author={Orangi, Behzad Mohammadi and Ghanati, Hadi Abbaszadeh and Basereh, Aref and Hesar, Narmin Ghani Zadeh and Jones, Paul A},
  journal={Scientific Reports},
  volume={15},
  number={1},
  pages={32206},
  year={2025},
  publisher={Nature Publishing Group UK London}
}

@article{moro2022markerless,
  title={Markerless vs. marker-based gait analysis: A proof of concept study},
  author={Moro, Matteo and Marchesi, Giorgia and Hesse, Filip and Odone, Francesca and Casadio, Maura},
  journal={Sensors},
  volume={22},
  number={5},
  pages={2011},
  year={2022},
  publisher={MDPI}
}

@article{carriere2025can,
  title={Can Machine Learning Enhance Computer Vision-Predicted Wrist Kinematics Determined from a Low-Cost Motion Capture System?},
  author={Carriere, Joel and Oliver, Michele L and Hamilton-Wright, Andrew and Young, Calvin and Gordon, Karen D},
  journal={Applied Sciences},
  volume={15},
  number={7},
  pages={3552},
  year={2025},
  publisher={MDPI}
}

@article{coleman2024gaining,
  title={Gaining more from doing less? The effects of a one-week deload period during supervised resistance training on muscular adaptations},
  author={Coleman, Max and Burke, Ryan and Augustin, Francesca and Pi{\~n}ero, Alec and Maldonado, Jaime and Fisher, James P and Israetel, Michael and Korakakis, Patroklos Androulakis and Swinton, Paul and Oberlin, Douglas and others},
  journal={PeerJ},
  volume={12},
  pages={e16777},
  year={2024},
  publisher={PeerJ Inc.}
}

@article{baayen2008mixed,
  title={Mixed-effects modeling with crossed random effects for subjects and items},
  author={Baayen, R Harald and Davidson, Douglas J and Bates, Douglas M},
  journal={Journal of memory and language},
  volume={59},
  number={4},
  pages={390--412},
  year={2008},
  publisher={Elsevier}
}

@misc{jefitExerciseDatabase,
	author = {JEFIT},
	title = {{E}xercise {D}atabase - {J}{E}{F}{I}{T} --- jefit.com},
	howpublished = {\url{https://www.jefit.com/exercises}},
	year = {2026},
	note = {[Accessed 26-01-2026]},
}

@article{pallares2021effects,
  title={Effects of range of motion on resistance training adaptations: A systematic review and meta-analysis},
  author={Pallar{\'e}s, Jes{\'u}s G and Hern{\'a}ndez-Belmonte, Alejandro and Mart{\'\i}nez-Cava, Alejandro and Vetrovsky, Tomas and Steffl, Michal and Courel-Ib{\'a}{\~n}ez, Javier},
  journal={Scandinavian journal of medicine \& science in sports},
  volume={31},
  number={10},
  pages={1866--1881},
  year={2021},
  publisher={Wiley Online Library}
}

@article{riemann2002sensorimotor,
  title={The sensorimotor system, part I: the physiologic basis of functional joint stability},
  author={Riemann, Bryan L and Lephart, Scott M},
  journal={Journal of athletic training},
  volume={37},
  number={1},
  pages={71},
  year={2002}
}

@article{scataglini2024accuracy,
  title={Accuracy, validity, and reliability of markerless camera-based 3D motion capture systems versus marker-based 3D motion capture systems in gait analysis: a systematic review and meta-analysis},
  author={Scataglini, Sofia and Abts, Eveline and Van Bocxlaer, Cas and Van den Bussche, Maxime and Meletani, Sara and Truijen, Steven},
  journal={Sensors},
  volume={24},
  number={11},
  pages={3686},
  year={2024},
  publisher={MDPI}
}

@article{colyer2018review,
  title={A review of the evolution of vision-based motion analysis and the integration of advanced computer vision methods towards developing a markerless system},
  author={Colyer, Steffi L and Evans, Murray and Cosker, Darren P and Salo, Aki IT},
  journal={Sports medicine-open},
  volume={4},
  number={1},
  pages={24},
  year={2018},
  publisher={Springer}
}

@article{jiang2023rtmpose,
  title={{RTMpose: Real-time multi-person pose estimation based on MMpose}},
  author={Jiang, Tao and Lu, Peng and Zhang, Li and Ma, Ningsheng and Han, Rui and Lyu, Chengqi and Li, Yining and Chen, Kai},
  journal={arXiv preprint arXiv:2303.07399},
  year={2023}
}

@article{redmon2018yolov3,
  title={Yolov3: An incremental improvement},
  author={Redmon, Joseph and Farhadi, Ali},
  journal={arXiv preprint arXiv:1804.02767},
  year={2018}
}

@article{burd2012muscle,
  title={Muscle time under tension during resistance exercise stimulates differential muscle protein sub-fractional synthetic responses in men},
  author={Burd, Nicholas A and Andrews, Richard J and West, Daniel WD and Little, Jonathan P and Cochran, Andrew JR and Hector, Amy J and Cashaback, Joshua GA and Gibala, Martin J and Potvin, James R and Baker, Steven K and others},
  journal={The Journal of physiology},
  volume={590},
  number={2},
  pages={351--362},
  year={2012},
  publisher={Wiley Online Library}
}

@inproceedings{yang2023effective,
  title={Effective whole-body pose estimation with two-stages distillation},
  author={Yang, Zhendong and Zeng, Ailing and Yuan, Chun and Li, Yu},
  booktitle={Proceedings of the IEEE/CVF International Conference on Computer Vision},
  pages={4210--4220},
  year={2023}
}

@misc{lu2023rtmo,
      title={{RTMO}: Towards High-Performance One-Stage Real-Time Multi-Person Pose Estimation},
      author={Peng Lu and Tao Jiang and Yining Li and Xiangtai Li and Kai Chen and Wenming Yang},
      year={2023},
      eprint={2312.07526},
      archivePrefix={arXiv},
      primaryClass={cs.CV}
}

@article{paszke2019pytorch,
  title={Pytorch: An imperative style, high-performance deep learning library},
  author={Paszke, Adam and Gross, Sam and Massa, Francisco and Lerer, Adam and Bradbury, James and Chanan, Gregory and Killeen, Trevor and Lin, Zeming and Gimelshein, Natalia and Antiga, Luca and others},
  journal={Advances in neural information processing systems},
  volume={32},
  year={2019}
}

@software{yolo26_ultralytics,
  author = {Glenn Jocher and Jing Qiu},
  title = {Ultralytics YOLO26},
  version = {26.0.0},
  year = {2026},
  url = {https://github.com/ultralytics/ultralytics},
  orcid = {0000-0001-5950-6979, 0000-0003-3783-7069},
  license = {AGPL-3.0}
}

@article{lu2018novel,
  title={Novel high-precision simulation technology for high-dynamics signal simulators based on piecewise hermite cubic interpolation},
  author={Lu, Shaozhong and Wang, Yongqing and Wu, Yunyun},
  journal={IEEE Transactions on Aerospace and Electronic Systems},
  volume={54},
  number={5},
  pages={2304--2317},
  year={2018},
  publisher={IEEE}
}

@article{yaro2023outlier,
  title={Outlier detection in time-series receive signal strength observation using Z-score method with Sn scale estimator for indoor localization},
  author={Yaro, Abdulmalik Shehu and Maly, Filip and Prazak, Pavel},
  journal={Applied Sciences},
  volume={13},
  number={6},
  pages={3900},
  year={2023},
  publisher={MDPI}
}

@article{cerna2000fundamentals,
  title={The fundamentals of FFT-based signal analysis and measurement},
  author={Cerna, Michael and Harvey, Audrey F},
  journal={National Instruments, Junho},
  volume={54},
  year={2000}
}

@inproceedings{akiba2019optuna,
  title={Optuna: A next-generation hyperparameter optimization framework},
  author={Akiba, Takuya and Sano, Shotaro and Yanase, Toshihiko and Ohta, Takeru and Koyama, Masanori},
  booktitle={Proceedings of the 25th ACM SIGKDD international conference on knowledge discovery \& data mining},
  pages={2623--2631},
  year={2019}
}

@article{balci2025reliability,
  title={Reliability assessment of markerless technologies in biomechanical motion analysis: a performance comparison},
  author={Balci, Ibrahim Cem and Sayin, Irem and Salturk, Serkan and Gursoy, Rana and Ozsoy, Umut and Dogru, Husnu Caglar and Akca, Gokhan and Eraslan, Ali and Sahin, Onurcan and Demircali, Ali Anil and others},
  journal={Frontiers in Sports and Active Living},
  volume={7},
  pages={1712332},
  year={2025},
  publisher={Frontiers Media SA}
}

@article{scataglini2025systematic,
  title={A systematic review of the accuracy, validity, and reliability of markerless versus marker camera-based 3D motion capture for industrial ergonomic risk analysis},
  author={Scataglini, Sofia and Fontinovo, Eugenia and Khafaga, Nouran and Khan, Muhammad Ubaidullah and Faizan Khan, Muhammad and Truijen, Steven},
  journal={Sensors},
  volume={25},
  number={17},
  pages={5513},
  year={2025},
  publisher={MDPI}
}

@article{williams2026effects,
  title={Effects of variable pose input on neural network model performance for classifying basketball player activity},
  author={Williams, Jadal and Poux-Guillaume, Tim and Hosoi, AE},
  journal={Journal of Sports Analytics},
  volume={12},
  pages={22150218261419241},
  year={2026},
  publisher={SAGE Publications Sage UK: London, England}
}

\end{document}